%% file: v850cen.tex
\documentclass{aa}  

\usepackage{graphicx}
\usepackage[normalem]{ulem}
\usepackage{pdfpages}
\usepackage{subfig}
\usepackage{float}
\usepackage{hyperref}
\hypersetup{
    colorlinks,
    linkcolor={red!50!black},
    citecolor={blue!50!black},
    urlcolor={blue!80!black}
}
\usepackage{amsmath}
\usepackage{xr} 
\usepackage{txfonts}
\usepackage{natbib}
\usepackage{inputenc}
\usepackage{longtable, pdflscape, multirow, booktabs}
\usepackage{balance}
\newcounter{subeq}
\newcommand{\stags}{\addtocounter{equation}{+1}\setcounter{subeq}{0}}
\newcommand{\stag}{\addtocounter{subeq}{1}\theequation\alph{subeq}}	

\def \inte {\emph{INTEGRAL}}

\def \gx {\mbox{GX~$304$-$1$}}
\def \vc {\mbox{V$850$~Cen}}

\begin{document}

   \title{Optical spectroscopy of the Be/X-ray Binary \vc/\gx~during faint X-ray periodical activity}

   \titlerunning{Optical observations of \vc}

   \author{C. Malacaria\inst{\ref{inst1}}
          \and
          W. Kollatschny\inst{\ref{inst2}}
          \and
          E. Whelan\inst{\ref{inst3}}
          \and
          A. Santangelo\inst{\ref{inst1},\ref{inst4}}
	  \and
	  D. Klochkov\inst{\ref{inst1}}
          \and
	  V. McBride\inst{\ref{inst5},\ref{inst6}}
          \and
          L. Ducci\inst{\ref{inst1},\ref{inst7}}
          }

   \institute{Institut f\"{u}r Astronomie und Astrophysik, Sand 1, 72076 T\"{u}bingen, Germany\\
              \email{malacaria@astro.uni-tuebingen.de}\label{inst1}
    \and Institut f\"{u}r Astrophysik, Universit\"{a}t G\"{o}ttingen, Friedrich-Hund Platz 1, 37077, G\"{o}ttingen, Germany\label{inst2}
    \and Maynooth University Department of Experimental Physics, National University of Ireland Maynooth, Maynooth Co. Kildare, Ireland\label{inst3}
    \and Institute of High Energy Physics, Chinese Academy of Sciences, Beijing 100049, China\label{inst4}
    \and Department of Astronomy, University of Cape Town, Private Bag X3, Rondebosch, 7701, South Africa\label{inst5}
    \and South African Astronomical Observatory, PO Box 9, Observatory, 7935, South Africa\label{inst6}
    \and ISDC Data Center for Astrophysics, Université de Genève, 16 chemin d'Écogia, 1290, Versoix, Switzerland\label{inst7}}

   \date{\today}

\abstract
{Be/X-ray binaries are the most populous class of High Mass X-ray Binaries. 
Their X-ray duty cycle is tightly related to the optical companion wind activity, 
which in turn can be studied through optical spectroscopical dedicated observations.} 
{We study optical spectral features of the Be circumstellar disk to test their 
long-term variability and their relation with the X-ray activity.
Special attention has been given to the H$\alpha$ emission line, 
that is one of the best tracers of the disk conditions.} 
{We obtained optical broadband medium resolution spectra from a dedicated campaign 
with the Anglo-Australian Telescope and the Southern African Large Telescope in $2014-2015$. 
Data span over one entire binary orbit, and cover both X-ray quiescent and moderately active periods.
We used Balmer emission lines to follow the evolution of the circumstellar disk.} 
{We observe prominent spectral features, like double-peaked H$\alpha$ and H$\beta$ emission lines. 
The H$\alpha$ V/R ratio significantly changes over a time scale of about one year.
Our observations are consistent with a system observed at a large 
inclination angle ($i\gtrsim60^{\circ}$). 
The derived circumstellar disk size shows that 
the disk evolves from a configuration that prevents accretion onto the neutron star,
to one that allows only moderate accretion.
This is in agreement with the contemporary observed X-ray activity.
Our results are interpreted within the context of inefficient tidal truncation 
of the circumstellar disk, as expected for this source's binary configuration.
We derived the H$\beta$-emitting region size, which results about half of the corresponding H$\alpha$-emitting disk,
and constrain the luminosity class of \vc~as III--V, consistent with the previously proposed class.}
{}


   \maketitle
%

\section{Introduction}
\input{Chapters/1_Introduction.tex}
\label{sec:intro}

\section{Observations and data reduction}
\input{Chapters/2_Data.tex}
\label{sec:observation}

\section{Results}
\input{Chapters/3_Results.tex}
\label{sec:results}




\section{Conclusions}
\input{Chapters/4_Conclusions.tex}
\label{sec:conclusions}

\begin{acknowledgements}
Some of the data used in this paper were acquired through the Australian Astronomical Observatory, via program UC$205$,
while other observations were obtained with the Southern African Large Telescope (SALT/HRS).
This work is supported by the \textsl{Bundesministerium f\"{u}r Wirtschaft und Technologie} 
through the \textsl{Deutsches Zentrum f\"{u}r Luft- und Raumfahrt e.V. (DLR)} 
under the grants FKZ 50 OR 1204. LD acknowledges the grant FKZ 50 OG 1602.
WK acknowledges the grant DFG Ko 857/32-2.
\end{acknowledgements}


\balance

\bibliographystyle{aa} 
\bibliography{v850cen}

\end{document}

%% file: Chapters/1_Introduction.tex
V850~Cen/\gx~is a Be/X-ray Binary system (BeXRB) composed of a B2~Vne optical companion 
(V850~Cen, m$_R\sim12.6$) and a pulsating neutron star (NS, \citealt{McClintock+77}),
located at a distance of $2.4\pm0.5\,$kpc \citep{Parkes+80}.
The first X-ray emission from this system has been detected in $1967$ during 
a balloon observation \citep{Hewish+68, Lewin+68}, while the optical 
counterpart has been discovered only ten years later, using data from 
the Anglo-Australian Telescope (AAT) by \citet{Mason+78}.

BeXRBs are characterized by luminous X-ray outbursts powered by the accretion of matter onto the NS.
The accreted matter is supplied by the Be star, whose stellar wind is expelled
under the form of a circumstellar equatorial disk,
which can be extended up to many stellar radii \citep{Rivinius+Carciofi13}.
Typically, X-ray (Type I) outbursts in BeXRBs take place at each periastron passage,
where the disk is closer to the NS orbit and the gravitational influence of the NS is stronger.
They are therefore characterized by a periodical recurrence identical with the orbital period.

\begin{figure*}[!htbp]
\centering
\includegraphics[width=\linewidth]{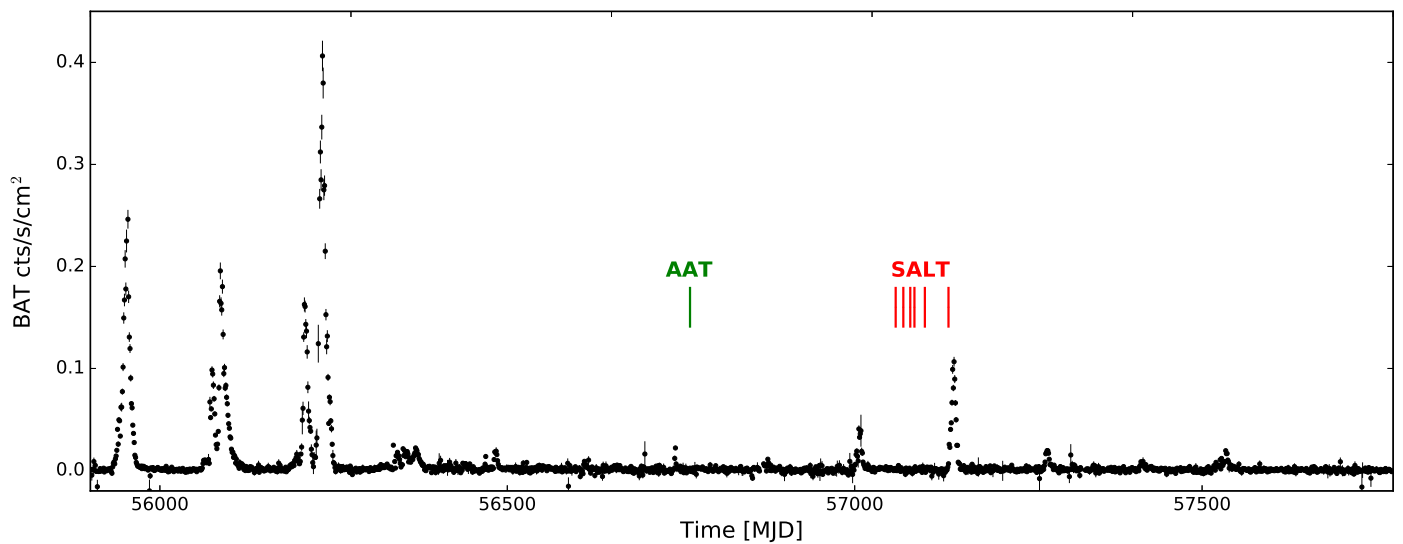}   
 \caption{Swift/BAT [$15-50\,$keV] light curve of the long term \gx~X-ray activity.
	Superimposed are the AAT/UCLES ($2014$ - green bar) and SALT/HRS ($2015$ - red bars) observations.}
\label{fig:opt_over}
\end{figure*}

\input{Tables/V850observation_log.tex}

The circumstellar decretion disk is a geometrically thin, high-density plasma 
rotating at a nearly Keplerian speed, and it is observationally determined by 
infrared (IR) excess and by Balmer emission lines in the optical band, 
especially the H$\alpha$ emission line \citep{Porter+03}.
In fact, spectral parameters of the H$\alpha$ line (such as the Equivalent Width, 
the Full Width at Half Maximum, and others) provide quantitative information on the 
circumstellar disk structure, its size and density distribution 
(see e.g. \citealt{Grundstrom+Glies06}, and references therein).

In certain cases, circumstellar disks in BeXRBs are truncated by the gravitational 
influence of the NS (\citealt{Negueruela+Okazaki01, Monageng+17}, and references therein), 
thus becoming optically thick at IR wavelengths and subject to radiation-driven instabilities.
Under the influence of such instabilities, the disk becomes warped, tilted, 
and precessing \citep{Porter98}.
Once deformed, the circumstellar disk can transfer matter onto the NS at different 
orbital phases than the periastron, thus powering giant (Type II) outbursts or shutting 
down the periodical ones (\citealt{Okazaki+13, Martin+14}, and references therein).
Therefore, the X-ray activity and the Be-disk status are tightly 
entwined and need to be studied altogether.

The X-ray history of \gx~is studded with both active and quiescent periods.
After its first X-ray detection, the source was regularly outbursting
with a ${\sim}132.5\,$d periodicity \citep{Priedhorsky+Terrell83}, which was interpreted as the orbital period.
The periodical X-ray activity lasted until around $1984$ \citep{Pietsch+86}, 
when the source entered a long period with no detectable X-ray emission, 
i.e. a quiescent phase.
The source resumed its activity only $28\,$years later, in June $2008$, 
when it was detected in the hard X-ray band by \inte~\citep{Manousakis+08}.
Since then, \gx~resumed its periodical outburst activity, 
with typical peak luminosities of $10^{36-37}\,$erg/s.
The new active period lasted for ${\sim}6\,$years, and its last outburst was a giant one,
characterized by an X-ray luminosity of $\sim2\times10^{37}\,$erg/s and 
a double-peaked lightcurve whose peaks were centered far from the periastron.
Afterwards, the source turned into another quiescent period,
showing only sporadic weak outbursts at periastron passages (see, e.g., \citealt{Nakajima+15}).

Despite \gx~being a luminous X-ray transient that has furnished a wealth of information on the physics
of accreting pulsars \citep{Devasia+11, Yamamoto+11, Klochkov+12, Malacaria+15, Jaisawal+16, Rothschild+17},
optical studies of \vc~are rare. 
However, early observations of \vc~during the $28\,$years X-ray quiescent period 
showed that the optical spectrum initially exhibited strong double-peaked 
H$\alpha$ line that progressively faded out, eventually leaving only 
the absorption feature \citep{Corbet+86, Haefner88}.
The lack of H$\alpha$ line from \vc~reflects the depletion of the Be-disk,
which turned off the X-ray source \citep{Pietsch+86}.

Here we present a spectroscopic study of \vc~in the wavelength range $4500$ to $8900\,\AA$.
Observations were obtained between April $2014$ and April $2015$, when the X-ray source 
was showing only moderate activity.
The results presented here are focused on a spectral analysis of the H$\alpha$ line, 
with the aim of probing the Be circumstellar decretion disk, 
and thus the connection between the disk and the X-ray activity of the NS.

%% file: Tables/V850observation_log.tex
\begin{table*}
\caption{Journal of V850~Cen optical observations with both AAT/UCLES and SALT/HRS.}\label{table:log}
{\small
\begin{center}
\begin{tabular}{ccccc}
\hline\hline
Program \& Block ID          & Seeing (arcsec) & Date \& Pointing time & Exposure [ks] & Filter \\[0.5ex]

\hline

 AAT UCLES + EEV2 UC205      & 1.50            & 2014-04-16 12:35  [UT]    & $2\times1200$  & -- \\
 2014-2-SCI-077 P1 HRS 32843 & 1.36            & 2015-02-06 01:23  [SAST]    & 940  		& SDSSr-S  \\
 2014-2-SCI-077 P1 HRS 32840 & 1.20            & 2015-02-17 01:32  [SAST]    & 940  		& SDSSi-S1 \\
 2014-2-SCI-077 P2 HRS 32841 & 1.90            & 2015-02-27 01:04  [SAST]    & 1034 		& SDSSg-S1 \\
 2014-2-SCI-077 P2 HRS 32842 & 1.60            & 2015-03-05 23:20  [SAST]    & 1034 		& SDSSi-S  \\
 2014-2-SCI-077 P3 HRS 32844 & 1.36            & 2015-03-20 03:02  [SAST]    & 940  		& SDSSr-S1 \\
 2014-2-SCI-077 P2 HRS 32839 & 1.54            & 2015-04-23 21:42  [SAST]    & 1034 		& SDSSi-S1 \\

\hline
\end{tabular}
\end{center}
}
\end{table*}

%% file: Chapters/2_Data.tex
\label{sec:data}

Optical spectroscopic observations of \vc~have been performed 
with two ground-based telescopes: 
the $3.9\,$m Anglo Australian Telescope (AAT) and the $10\,$m 
Southern African Large Telescope (SALT).
A log of the observations is given in Table~\ref{table:log}, 
while Fig.~\ref{fig:opt_over} shows the optical observations 
superimposed to the X-ray long-term light curve of \gx.

\input{Tables/Line_parameters.tex}
The first observation has been obtained on $2014$, April 16th, at the 
AAT with the University College London Echelle Spectrograph (UCLES, \citealt{Horton+12}).
UCLES is a high-resolution spectrograph (R$~>40000$) sensible 
in the range $4500$-$8900\,\AA$.
The source was observed with a $2\times20\,$min exposure and a 
$1\farcs5$ slit width while the seeing was $1\farcs50$.
Both runs were carried out with the $31\,$line/mm cross-dispersing 
grating and the EEV2 CCD (with a central wavelength of $5450\,\AA$).
The chip was read out in "normal" mode, with a readout noise of $3.9\,$electrons.

The second set of observations was carried out at the SALT
with the High Resolution Spectrograph (HRS, \citealt{Tyas12}).
SALT/HRS is a dual-beam (blue arm $370-555\,$nm and red arm $555-890\,$nm) 
fibre-fed echelle spectrograph, with a single $2k\times4k$ CCD to capture 
all the blue orders, while a $4k\times4k$ detector is used for the red orders.
HRS was always set up in the Medium Resolution Mode (R$\sim40000$),
whose readout speed is $400\,$kHz for both arms and the slit width is $0\farcs7$.
The CCDs binning was the standard $1\times1$ in the spatial direction,
that is optimized for the spectral resolution.
Due to problems with the CCD in the blue arm science frame 
(blue spectra suffer from instrumental contamination and are complicated by 
multiple read-out amplifiers), here we have only used SALT red arm data.

All data have been reduced using standard \textit{IRAF}\footnote{Image Reduction 
and Analysis Facility: iraf.noao.edu} packages (version $2.16$), 
and all images have been both bias- and flatfield-corrected.
Wavelength calibration has been applied using arc (Thorium-Argon) lamp spectra.
Flux calibration was not applied, since the spectra were used to measure
spectral parameters of emission features and to locate the peaks of the double-peaked emission lines.

%% file: Tables/Line_parameters.tex
\begin{table*}[!t]
\caption{Spectral parameters of the H$\alpha$ emission line and derived disk radii.}\label{table:line_params}
{\small
\begin{center}
\begin{tabular}{cccccccccc}
\hline
\midrule
MJD ($\phi_{orb}$)         & EW      & FWHM 		       & $\Delta$V      & V/R 		& \multicolumn{4}{c}{${V}\sin i$ [km/s]} & R$_{disk}$\\[0.5ex]
             & [\AA]   & [\AA] (km/s)	       &  [km/s]        &     		&    Eq.~$1$ & Eq.~$2$a & Eq.~$2$b &Eq.~$3$ & [R$_\star$]\\[0.5ex]
\midrule
 56763 (0.12) & $-14.1\pm0.7$ & $7.8\pm0.1\, (367\pm3)$ & $252\pm3$  & $0.22\pm0.03$ & $226$ & $466$ & $323$ & $218$ & $5.9\pm0.3$  \\
 57059 (0.36) & $-36.4\pm1.2$ & $7.4\pm0.1\, (342\pm3)$ & $185\pm6$  & $0.83\pm0.02$ & $209$ & $463$ & $325$ & $223$ & $11.1\pm3.5$ \\
 57070 (0.44) & $-32.3\pm0.9$ & $7.1\pm0.1\, (326\pm3)$ & $180\pm3$  & $0.75\pm0.01$ & $197$ & $434$ & $304$ & $210$ & $10.2\pm1.8$ \\
 57080 (0.52) & $-33.0\pm1.2$ & $7.3\pm0.2\, (337\pm4)$ & $189\pm3$  & $0.73\pm0.02$ & $205$ & $458$ & $321$ & $218$ & $10.4\pm2.1$ \\
 57086 (0.56) & $-39.0\pm1.3$ & $7.3\pm0.2\, (334\pm3)$ & $191\pm3$  & $0.73\pm0.03$ & $203$ & $489$ & $343$ & $220$ & $11.6\pm2.8$ \\
 57101 (0.68) & $-32.8\pm1.5$ & $7.6\pm0.2\, (354\pm4)$ & $200\pm3$  & $0.84\pm0.01$ & $217$ & $484$ & $339$ & $229$ & $10.3\pm2.3$ \\
 57135 (0.94) & $-27.0\pm1.3$ & $8.1\pm0.4\, (366\pm4)$ & $224\pm9$  & $0.75\pm0.05$ & $226$ & $507$ & $356$ & $232$ & $9.1\pm2.1$  \\

\midrule
\end{tabular}
\end{center}
}
\end{table*}

%% file: Chapters/3_Results.tex
\externaldocument{1_Introduction.tex}
\externaldocument{2_Data.tex}

\subsection{The rotational velocity}\label{subsec:rot_vel}

It is known that Be stars are rapid rotators, and their rotational velocity
can be measured using emission or absorption lines parameters. 
\citet{Steele+99} provide the relations between the projected rotational velocity 
${V}\sin i$ (where $i$ is the inclination angle under which the observer sees the disk plane)
and the Full Width at Half Maximum of He~I lines.
Those relations have been largely employed in the literature
(see, e.g. \citealt{Rajoelimanana+17, Reig+10, Kiziloglu+07, Reig+04}).
However, He~I lines lie in the blue part of the stellar spectrum, 
which is not available in our case (see Sect.~\ref{sec:data}).
On the other hand, \citet{Hanuschik89} derived the relations between ${V}\sin i$ 
and the spectral parameters of the H$\alpha$ emission line, i.e. the peak separation 
${\Delta}V$ (in case of double-peaked lines), the Equivalent Width EW, and the Full Width at Half Maximum FWHM.
Those relations are described in the following equations, where the ${V}\sin i$ is given 
as a function of the FWHM:

\begin{equation}\label{eq:one}
FWHM(H\alpha) = 1.4{V}\sin i + 50\,km\,s^{-1}\, ,
\end{equation}
\noindent
as a function of ${\Delta}V$ and EW:

\begin{align}
\stags
\log\,\left(\frac{\Delta\,V}{2{V}\sin i}\right) = -0.32\log\,EW(H\alpha)-0.20\tag{\stag}\, ,     \label{eq:two}
\end{align}
\noindent
or as a function of FWHM and EW:

\begin{equation}\label{eq:three}
\log\,\left(\frac{FWHM}{2{V}\sin i}\right) = -0.1\log\,EW(H\alpha)+0.04\, ,
\end{equation}

\noindent
with $\Delta$V, $FWHM$, and ${V}\sin i$ measured in km\,s$^{-1}$, while $EW$ in \AA 
(where the absolute value of EW is considered as the argument of the logarithm).

Because the fit parameters are reported without errors in \citet{Hanuschik89},
a realistic error calculation on the derived ${V}\sin i$ values is not possible.
Following \citet{vanbelle2008} for an estimation of the error on ${V}\sin i$
we considered the range shown by the data points in Fig.~s~$1$, $4$ and $8$ in \citet{Hanuschik89},
from which Eq.~s~\eqref{eq:one}, \eqref{eq:three} and \eqref{eq:two}, respectively, have been obtained.
The resulting errors on ${V}\sin i$ are $50$, $40$, and $110$ km s$^{-1}$ for Eq.~s~\eqref{eq:one},
\eqref{eq:three}, and \eqref{eq:two}, respectively.
\begin{figure}[!t]
\centering
\includegraphics[width=\linewidth, trim = 1cm 1cm 1cm 1cm, clip]{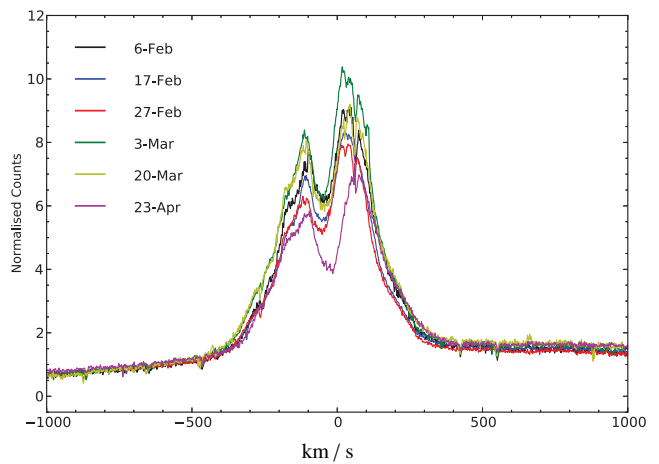} %
\vspace*{0.1cm}\hspace*{0.0cm}\textcolor{black}{\tiny km / s}
\caption{Evolution of the H$\alpha$ emission line profile, as observed by SALT/HRS 
	during one entire orbital period at the beginning of $2015$.}
\label{fig:Halpha}
\end{figure}

We calculated EWs for double-peaked profiles by direct integration of the flux 
across the feature using the \textit{splot} task of \textit{IRAF}.
The same task has been used to obtain FWHMs by fitting a single Gaussian profile 
over the entire double-peaked feature, which fits well the overall line shape.
Furthermore, the double-peaked profiles have also been fitted using the deblending routine available in IRAF.
The two peaks have been modelled with two Voigt functions, 
which resulted to fit the line wings better than Gaussian functions.
This allowed us to obtain line centres and intensity of the two peaks above the continuum.
These, in turn, allowed us to calculate the so-called $V/R$ ratio, 
defined as the relative intensity of the blue (V) and red (R) peaks 
in the split profile of the line, and the peak separation ${\Delta}V$.
Following the methodology of \citet{Reig+10}, to derive all H$\alpha$ lines spectral parameters 
the fitting procedure has been iterated twelve times for each double-peaked feature.
Since the main source of uncertainty in the EW is due to the difficult definition
of the continuum, at each iteration a slightly different point of the continuum was sampled.
Final values and errors of line parameters have been calculated as the average and 
standard deviation over those iterated measures.

\begin{figure}[!t]
\includegraphics[width=\linewidth]{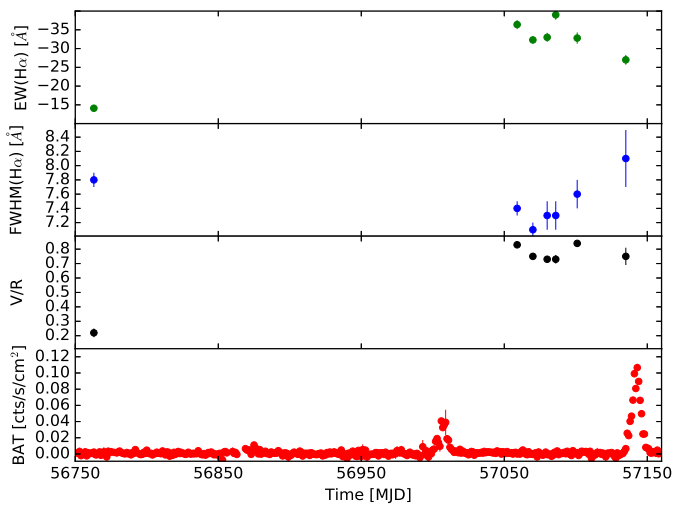}
 \caption{Variability of the H$\alpha$ line spectral parameters compared to the X-ray flux as a function of time.
	Panels show, from top to bottom, the Equivalent Width (EW), the Full Width Half Maximum (FWHM), 
	the V/R ratio, and the Swift/BAT X-ray count rate,
	while on the x-axis is reported the time in MJD.}
\label{fig:correlations1}
\end{figure}
The results are given in Table~\ref{table:line_params}.
Values of ${V}\sin i$ obtained by Eq.~\eqref{eq:one}, \eqref{eq:two}, 
\eqref{eq:three} show different nominal values (consistent within the large errors).
However, it is important to note that those relations are obtained for isolated Be stars, while
circumstellar disks of BeXRBs are about $1.5 - 2$ times denser than 
those of isolated Be stars \citep{Zamanov+01, Reig+16}.
Such an effect has been attributed to disk truncation, although those works find 
higher disk densities also for BeXRBs with long orbital periods, in which disk truncation 
is expected to be inefficient (see Sect.~\ref{subsec:truncation}).
\citet{Hanuschik+88} relates the disk density to the intercept of Eq.~\eqref{eq:two}
(see Eq.~$8$ of their work), while \citet{Monageng+17} find that
for a group of BeXRBs, the best-fit solution between log($\frac{\Delta\,V}{2{V}\sin i}$) vs log($\frac{-EW}{\AA}$) 
results into the new relation: \addtocounter{equation}{-2}

\begin{align}
\stags
\log\,\left(\frac{\Delta\,V}{2{V}\sin i}\right) = -0.33\log\,EW(H\alpha) - 0.03 \tag{\stag}  \addtocounter{subeq}{+1}     \label{eq:two_b}
\end{align}

\noindent
\addtocounter{equation}{-2} Eq.~\eqref{eq:two_b} returns an average
projected velocity of $\sim330\pm55\,$km\,s$^{-1}$.
Since this value has been obtained by using a relationship that is calibrated on BeXRBs, 
we will use this as a reference value in the following.

\begin{figure}[!t]
\includegraphics[width=\linewidth]{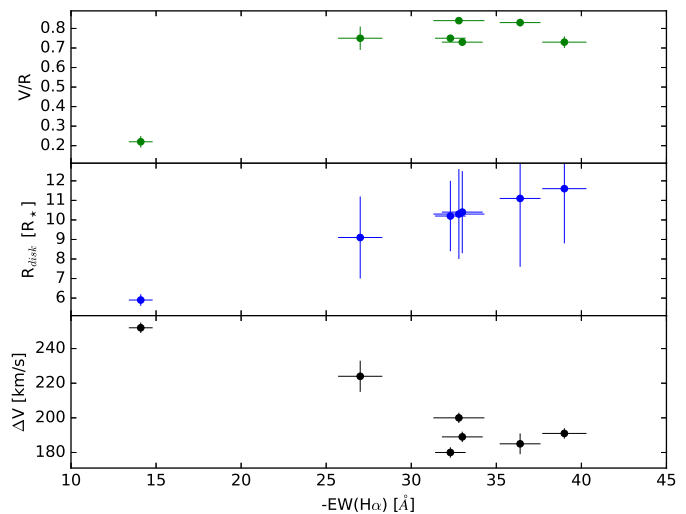}
 \caption{Correlations between the H$\alpha$ line-related parameters and the H$\alpha$ Equivalent Width.
	Panels show on the y-axis, from top to bottom, the $V/R$ ratio, 
	the circumstellar disk radius calculated from \addtocounter{equation}{+1} Eq.~\eqref{eq:radius}, 
	and the peak separation {$\Delta$}V,
	while on the x-axis is reported the EW absolute value.}
\label{fig:correlations2}
\end{figure}
Be stars are believed to rotate at $\sim60\%-80\%$ of their critical velocity 
\citep{Porter96, Rivinius+06}.
Earlier observations of \vc~H$\alpha$ line \citep{Parkes+80, Corbet+86}
detected a shell-type profile of the emission line, that is a double-peaked profile 
in which the central absorption extends below the stellar continuum flux.
According to the seminal model by \citet{Struve31} (yet largely accepted, 
see \citealt{Rivinius+Carciofi13} for a more recent review),
shell-type profiles imply an inclination angle of $i\sim90^{\circ}$ (i.e. an edge-on system).
During our campaign we do not observe shell-type emission lines (see Fig.~\ref{fig:Halpha}).
Therefore, based on a qualitative comparison with Fig.~$1$ of \citet{Rivinius+Carciofi13},
the profiles detected in this work support a large ($60^{\circ}{\lesssim}\,i\,{\lesssim}80^{\circ}$) 
inclination angle, yet not an edge-on view.
Later, more physical models \citep{Hummel94, Hanuschik96, Hummel+Hanuschik97} 
also support a large inclination angle.
However, we should note here that line profile shapes do not always give unambiguous 
information about the disk inclination angle \citep{Silaj+10, Quirrenbach+97}.
In any case, transient shell lines do not demand a tilt of the disk plane.
For example, shell profiles need a geometrically thick medium to form.
Therefore, the H$\alpha$ line profiles observed in the present work 
may simply be due to a decrease of the disk density, with respect to previous conditions.
Assuming typical values of the mass and radius of a B2 type star as 
M$_\star=9.9\,M_{\odot}$, R$_\star=5.4\,$R$_\odot$ (\citealt{Townsend+04, Pasinetti+00, Sugizaki+15}),
the critical velocity of rotation is 
\begin{equation}
V_{break}=\sqrt{\frac{GM_\star}{R_\star}}\approx475\,$km\,s$^{-1}
\end{equation}
With an inclination angle $i=90^{\circ}$ and ${V}\sin i=330\,$km\,s$^{-1}$,
the resulting critical fraction is $w=V/V_{break}\approx0.69$. 
On the other hand, assuming $w=0.8$ and ${V}\sin i=330\,$km\,s$^{-1}$, the resulting inclination angle is $i\sim60^{\circ}$,
which is still consistent with the inclination inferred by the H$\alpha$ line profile.
We therefore conclude that, at least during our observations, a more conservative range for 
the inclination angle of \vc~is $60^{\circ}{\lesssim}\,i\,{\lesssim}90^{\circ}$ and that
\vc~is likely rotating at $0.69{\lesssim}\,w\,{\lesssim}0.80$.

\subsection{The H$\alpha$ line evolution}

Table~\ref{table:line_params} summarizes the results of the spectral analysis, 
showing the evolution of H$\alpha$ spectral parameters during the observational campaign.
The evolution of the H$\alpha$ emission line profile is also shown in Fig.~\ref{fig:Halpha},
while a comparison of the H$\alpha$ spectral parameters with the X-ray flux is shown in Fig.~\ref{fig:correlations1}.
From the latter plot, it is clear that the EW and the V/R ratio show a significant increase 
from the observation in $2014$ to those in $2015$.
The EW doubled on a time scale of one year, while the V/R ratio increased of a factor of $4$ in the same period.
Despite the large variation in the V/R ratio, the dominant profile is always $V<R$.
The V/R variability is an important tracer of the disk activity.
Indeed, while the strength of the H$\alpha$ line (i.e. the EW) helps to trace size and density 
of the circumstellar disk (see Sect.~\ref{subsec:disk_size}), its morphology gives information 
about the degree of symmetry in the disk density, and in particular to possible global one-armed 
density oscillations (\citealt{Oktariani+Okazaki09}, and references therein).
Our observations show that the disk density is asymmetrically distributed,
likely due to a global density perturbation, and that the perturbation is still 
present at the smallest measured EW (i.e. at the smallest disk radius), although at a much lower degree.
However, when the disk grows the perturbation can develop further and the V/R ratio becomes higher.
The relative intensity between the V and R peaks of Be stars in BeXRBs is 
found to cyclically vary (i.e. from $V<R$ to $V>R$, and vice versa) 
with a typical period of about $5\,$years \citep{Negueruela+Okazaki01, Reig+16}.
The very low value of the V/R ratio observed in $2014$ possibly indicates that, before our observation
the V/R ratio was even lower, with a V$\approx$R period, preceeded by a $V>R$ phase.

The evolution of H$\alpha$ spectral parameters seems also to reflect the evolution 
of X-ray production from the NS, as shown in Fig.~\ref{fig:opt_over}.
The minimum observed value of EW is obtained during the AAT observation on April $2014$.
Around that period, the X-ray source stayed on a quiescent level during repeated periastron passages.
One year later the EW increased significantly while, 
at the same time, the X-ray source was showing moderate activity \citep{Nakajima+15}.
Therefore, higher X-ray fluxes are produced when the H$\alpha$ line is stronger,
that is when the size of the H$\alpha$-emitting region, i.e. the circumstellar disk, is larger,
thanks to the fact that larger disks get closer to the NS.
Immediately before and after our SALT observations in $2015$, the source shows moderate X-ray outbursts 
at the periastron passages \citep{Nakajima+15}, but the following outbursts are missing, 
showing minor and only sporadic periodical activity.
This likely indicates that during our observations the circumstellar disk evolved 
from $2014$ to $2015$ in such a way that the transfer of a moderate amount of matter 
to the accreting NS was favored, but the disk conditions subsequently changed leading to the opposite trend.
At the time of writing, the source has still not resumed its typical outburst X-ray intensity.
This may be due to major disruption of the circumstellar disk, 
or perhaps to its distortion, thus preventing efficient accretion onto the NS.
More quantitative considerations about the disk changes are discussed in the next Sections.

\subsection{Circumstellar Be disk size}\label{subsec:disk_size}

\addtocounter{equation}{3}
The peak separation ${\Delta}V$ is an important parameter that helps to trace 
the circumstellar disk conditions: ${\Delta}V$ is related to the disk size 
and to the inclination angle $i$ (see Eq.~\ref{eq:two_b}).
A small peak separation results from a combination of a large circumstellar disk 
and a small inclination angle such that the disk is seen by the observer as almost 
uncovered by the central star (i.e. a face-on system).
Conversely, a large peak separation results from a combination of a smaller disk 
and a large inclination angle such that the disk is heavily obscured by the central star (i.e. an edge-on system).
For rotationally dominated line profiles, the peak separation ${\Delta}V$ of H$\alpha$ emission lines 
can be used to estimate the size of the H$\alpha$-emitting region R$_{d}$, 
i.e. the circumstellar disk size\footnote{Other works report the equation for R$_d$ 
in another form, see e.g. \citet{Coe+06}. However, Eq.~\eqref{eq:radius} has two main advantages. 
First, it takes into account the (usually incorrect) assumption that the star is critically rotating and, second, 
it uses the ${V}\sin i$ quantity, which is obtained directly by observations.} \citep{Huang72}:
\begin{equation}\label{eq:radius}
R_{d} = \left(\frac{2{V}\sin i}{\Delta V}\right)^j\,\epsilon R_\star
\end{equation}
where $j=2$ for Keplerian rotation, R$_\star$ is the central star radius, and $\epsilon$ is a 
dimensionless parameter that takes into account several effects that 
would overestimate the disk radius, $\epsilon=0.9\pm0.1$ \citep{Zamanov+13}.
To calculate the disk radius, a stellar radius of R$_\star=5.4\,$R$_\odot$ has been assumed 
(see Sect.~\ref{subsec:rot_vel}), while ${V}\sin i$ has been derived by Eq.~\eqref{eq:two_b}.
Fig.~\ref{fig:correlations2} (middle panel) shows the disk radius R$_{d}$ 
as a function of the EW(H$\alpha$) (absolute value).
A positive correlation between the two parameters is found, 
in agreement with the expectation that the EW(H$\alpha$) reflects the size 
of the circumstellar disk \citep{Quirrenbach+97, Grundstrom+Glies06}.

Fig.~\ref{fig:correlations2} (bottom panel) shows the peak separation 
${\Delta}V$ 
as a function of the EW(H$\alpha$) (absolute value).
A clear anti-correlation is present between the two parameters, indicating that
as the disk radius (i.e. the EW) grows, the peak separation decreases.
For strong EWs, the anti-correlation can be extrapolated down to a point where
the two peaks merge (${\Delta}V = 0$), and a single peaked H$\alpha$ line emerges.
For a number of BeXRBs, this occurs when the EW(H$\alpha$) becomes
stronger than ${\sim}-15\,\AA$ \citep{Reig+16}.
However, our observations shows that ${\Delta}V$ is still large at large
disk radii ($\left|EW\right|>15\,\AA$).
This may be due to the large inclination angle $i$, which makes the disk
to be seen almost edge-on.
Indeed, if $i\sim90^\circ$, then it is difficult for the disk to emerge out of
the optical companion obscuration, even at large radii.
Also, results by \citet{Silaj+10} indicate that broadened double-peaked H$\alpha$ profiles
(thus large ${\Delta}V$ values) are expected at large inclination angles (see Fig.~$5$ in their work).
On the other hand, as \citet{Reig+16} point out, the EW limit value is expected 
to depend on the spectral resolution which, in their case, is generally much lower
than ours (their spectra have typically R$~<2500$).

\input{Tables/Beta.tex}
To better understand the accretion dynamics, it is also interesting to compare 
the circumstellar disk size with the NS distance from \vc.
Assuming R$_\star=5.4\,$R$_\odot$, with ${V}\sin i$ and ${\Delta}V$ as 
observed with AAT ($294$ and $252\,$km\,s$^{-1}$, respectively, see Table~\ref{table:line_params}), 
a disk radius of $R_{d}\sim2.2\times10^{10}\,m\,\sim5.9\,$R$_\star$ is obtained.
On the other hand, assuming an orbital eccentricity of $e\sim0.5$ and a semi-major axis of
${a_x}\sim500\,$lt-s (for an inclination angle $i=90^\circ$, \citealt{Sugizaki+15}), 
the obtained periastron distance is $a_{peri}=a(1-e)\sim7.5\times10^{10}\,$m\,$\sim19\,$R$_\star$.
Therefore, when the NS star approaches the periastron around the time of our AAT observation,
the circumstellar matter is still within the Roche-lobe of \vc~
(R$_{RL}\sim9\,$R$_\star$, calculated using the approximation by \citealt{Eggleton83}), 
and the transfer of matter onto the NS is completely inhibited.
Indeed, the periodical X-ray outbursts around that period are missing (see Fig.~\ref{fig:opt_over}).
However, according to our SALT observations on $2015$, 
during that epoch the disk radius grew up to $R_{d}\sim9-11\,$R$_\star$
and this was enough to allow the transfer of matter onto the NS,
as reflected by the (weak) X-ray intensity of periodical outbursts during $2015$.
This is in agreement with the X-ray activity evolution observed 
at the two different epochs (see Fig.~\ref{fig:opt_over}).

\subsection{Inefficient tidal truncation}\label{subsec:truncation}

For closer binary orbits, the gravitational influence of the NS onto the circumstellar disk 
leads to resonances between the disk and the NS orbital period \citep{Okazaki+Negueruela01, Haigh+04, Coe+06}.
Such resonances tend to truncate the disk at specific radii, 
which can be calculated as (see also \citealt{Grundstrom+07b}):
\begin{equation}
R_n^{3/2} = \frac{\left(G M_\star\right)^{1/2}}{2\pi n}P_{orb}
\end{equation}
\noindent
where $G$ is the gravitational constant, M$_\star$ is the central star mass, 
P$_{orb}$ is the orbital period, and $n$ is the integer number of disk rotational periods per one orbital period.
However, \citet{Okazaki+Negueruela01} proposed that in high-eccentricity binary systems, 
such as \gx, disk truncation due to resonant effects is not efficient, 
and that the circumstellar disk radius is only limited by the closest approach 
of the NS to the companion, i.e. by the periastron distance.
Assuming P$_{orb}=132.189\,$d and M$_\star=9.9\,$M$_\odot$ \citep{Sugizaki+15},
the maximum observed disk radius (R$_d{\sim}11\,$R$_\star$, see Sect.~\ref{subsec:disk_size}) 
should be truncated by the $8$:$1$ resonance, which 
implies an unusually high value of $n$ for BeXRBs
(see \citealt{Monageng+17}, and references therein), and possibly indicates that the disk size 
of \vc~as observed in the present work did not reach its largest possible radial extension.
This possibility is further explored in the following.

We observe the strongest EW(H$\alpha$) during our SALT campaign at MJD~$57086$, that is $-39\,$\AA.
Among BeXRBs, the maximum observed EW is found to correlate with the orbital period 
of the binary system P$_{orb}$ \citep{Reig+97, Antoniou+09, Coe+Kirk15}
and with the periastron distance \citep{Reig+16}.
According to those authors, an orbital period of $\sim132\,$d (or, similarly, 
a periastron distance of $19\,$R$_\star$) would imply an EW of the H$\alpha$ emission 
line between $-30$ and $-40\,$\AA.
This is consistent with our observations, suggesting that the circumstellar disk 
of \vc~was approaching its maximum size around the time of observations with SALT.
However, our optical observations do not cover a typical X-ray outburst: the X-ray flux 
of periodical outbursts observed immediately before and after our SALT campaign 
is about half of the \gx~standard ones (see Fig.~\ref{fig:opt_over}). 
The lower X-ray flux is due to the lower mass accreted by the NS,
which in turn reflects the lack of disk material in the vicinity of the compact object.
Within this interpretation it is therefore possible that the maximum 
EW value observed in the present work does not represent the actual maximum size 
reachable by the circumstellar disk but rather a lower limit on the maximum size,
thus allowing \gx~to deviate from the above-mentioned correlations.
Since those correlations have been interpreted as evidence of disk truncation, 
the possibility of \gx~deviating from them is still consistent
with the model of inefficient tidal truncation in \gx~proposed by \citet{Okazaki+Negueruela01}.

Finally, we note that typical circumstellar disk radii in BeXRBs have $\leq6\,$R$_\star$ \citep{Reig+16}.
This value is smaller than that of circumstellar disks radii in isolated Be stars ($\gtrsim14\,$R$_\star$).
Also, the resulting upper limit value is independent from the donor star size.
The above considerations suggest that the maximum extent of the disk 
is a characteristic quantity of binary systems in which tidal truncation is at work.
On the other hand, \vc~develops a disk at least as large as $\sim11\,$R$_\star$.
Again, this result can be interpreted as a possible hint for inefficient tidal truncation in \gx.

\begin{figure}[!t]
\includegraphics[width=\linewidth, trim = 9cm 1cm 9cm 1cm, clip]{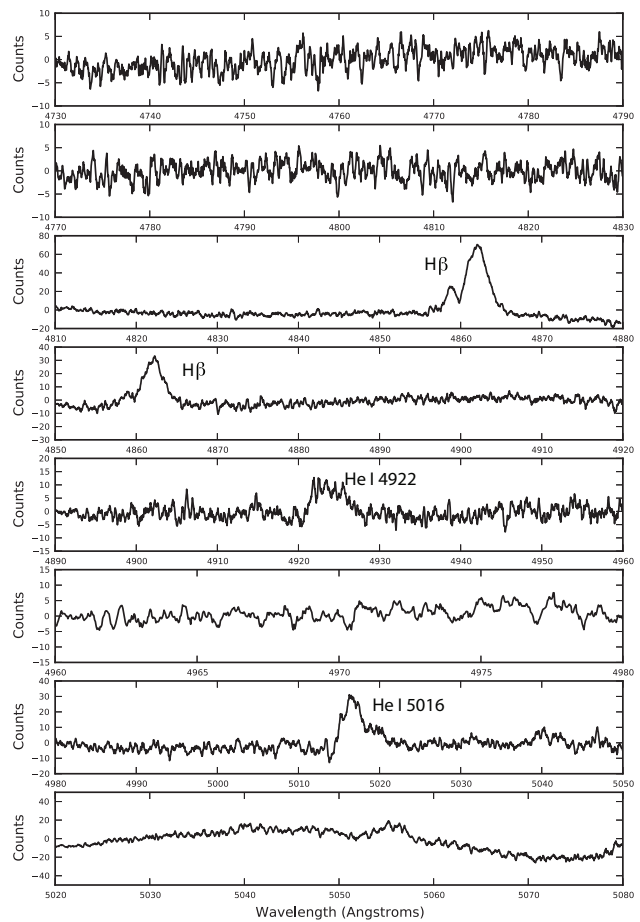} 
\caption{Selection of AAT/UCLES orders at bluer wavelengths.
	Highlighted are the prominent H$\beta$ line and He\,I~$4922, 5016$ features.}
\label{fig:AATpdf}
\end{figure}

\subsection{The H$\beta$ emission line}

Data from the AAT observation in $2014$ allowed us to measure the spectral parameters of the H$\beta$ emission line.
For this, we applied the same procedure used for the measurement of H$\alpha$ line parameters 
(see Sect.~\ref{subsec:rot_vel}).
The results are summarized in Table~\ref{table:beta_params}, while a selection of bluer 
AAT orders including the H$\beta$ emission line is shown in Fig.~\ref{fig:AATpdf}.
The very presence of an H$\beta$ emission line indicates that the disk is already dense,
even at its smallest measured radius.
The H$\beta$ emitting disk radius has been calculated according to Eq.~\eqref{eq:radius},
adopting a value of $j=1$ \citep{Mennickent+Vogt91} and results of about $3.2\,$R$_\star$.
This is in agreeement with observations of H$\beta$ emitting disks in other Be stars \citep{Mennickent91, Zamanov+16}.
However, we note that, contrary to other Be stars observations \citep{Hanuschik+88, Zamanov+16},
the H$\beta$ peak separation (and the overall line shape) results smaller than the H$\alpha$.
In the case of \citet{Hanuschik+88} the spectral resolution of the employed instrument 
is of the order of $R\sim10^5$, and this can be a possible reason for the observed discrepancy.
Moreover, as in the case for the H$\alpha$ line, the H$\beta$ also shows a $V<R$ line profile,
indicating that the asymmetry in the disk density extends down to smaller radii.

Finally, the EWs of H$\alpha$ and H$\beta$ lines can be compared to constrain the 
luminosity class of optical companions in High Mass X-ray Binaries \citep{Reig+96, Fabregat+96}.
Our results suggest a luminosity class III--V, which is consistent with the stellar classification proposed by \citet{Parkes+80}.

%% file: Tables/Beta.tex
\externaldocument{3_Results.tex}

\begin{table*}[!t]
\caption{Spectral parameters of the H$\beta$ emission line and derived disk parameters.}\label{table:beta_params}
{\small
\begin{center}
\begin{tabular}{cccccc}
\hline
\midrule
MJD          & EW      & FWHM 		        & ${\Delta}V$ & V/R 	 & R$_{disk}$ \\[0.5ex]
             & [\AA]   & [\AA] (km/s)	        &  [km/s]     &          & [R$_\star$] \\[0.5ex]
\midrule
 56763 & $-5.6\pm0.1$ & $3.8\pm0.1\, (235\pm3)$ & $180\pm2$  & $0.35\pm0.01$ & $3.2\pm0.6$ \\

\midrule
\end{tabular}
\end{center}

}
\end{table*}

%% file: Chapters/4_Conclusions.tex
We have analyzed optical spectra of \vc, the donor star of the BeXRB \gx.
Data were taken with the AAT/UCLES and SALT/HRS, spanning over one year and 
distributed along the orbital period of the binary system.
Our study has been focused on the Balmer emission lines, and the characterization of the circumstellar disk,
in order to get insights about the connection between the stellar wind structure of \vc~and the X-ray activity of \gx.
Our main results are summarized in the following:

\begin{itemize}

\item During all our observations, the H$\alpha$ line is always present in emission.
	Its profile is always double-peaked, although it shows strong variation on a time scale of ${\sim}1\,$year.

\item Our observations support a large inclination angle, i.e. $i\gtrsim60^{\circ}$, possibly an edge-on system.

\item The H$\alpha$-emitting circumstellar disk has a size that ranges from 
	${\sim}5.9\,R_\star$ in $2014$ to ${\sim}11\,R_\star$ one year later.
	With a Roche-lobe at the periastron of R$_{RL}{\sim}9\,$R$_\star$, 
	the X-ray intermittent activity of \gx~	is a direct consequence of the circumstellar disk evolution.

\item Several pieces of evidence point to the possibility of inefficient truncation of the circumstellar 
	disk by the orbiting NS, as expected from theoretical arguments for this source's binary configuration.
	However, no firm conclusion can be drawn by our data on this point.

\item Detection of the H$\beta$ emission line and its comparison to the H$\alpha$ line	
	allowed us to derive a circumstellar disk size of the H$\beta$ emitting region 
	that is about half of the H$\alpha$-emitting disk and to constrain the luminosity class 
	of the source as III--V, consistent with the previously proposed class.

\end{itemize}